\documentclass[prl,twocolumn,raggedbottom,superscriptaddress,nofootinbib]{revtex4-1}

\usepackage{graphicx}
\usepackage[colorlinks,linkcolor=blue,urlcolor=blue,citecolor=blue]{hyperref}
\usepackage{epsfig}
\usepackage{bm}

\begin{document}
\def\pks{Max-Planck Institute for the Physics of Complex Systems, N\"othnitzer Str.\ 38, 01187 Dresden, Germany}
\def\rice{Rice University, Department of Physics and Astronomy and Rice Center for Quantum Materials, Houston, Texas 77251, USA}
\title{Resonant Rydberg Dressing of Alkaline-Earth Atoms via Electromagnetically Induced Transparency}

\author{C. Gaul}
\affiliation{\pks}
\author{B. J. DeSalvo}
\affiliation{\rice}
\author{J. A. Aman}
\affiliation{\rice}
\author{F. B. Dunning}
\affiliation{\rice}
\author{T. C. Killian}
\affiliation{\rice}
\author{T. Pohl}
\affiliation{\pks}
\date{\today}

\begin{abstract}
We develop an approach to generate finite-range atomic interactions via optical Rydberg-state excitation and study the underlying excitation dynamics in theory and experiment. In contrast to previous work, the proposed scheme is based on resonant optical driving and the establishment of a dark state under conditions of electromagnetically induced transparency (EIT). Analyzing the driven dissipative dynamics of the atomic gas, we show that the interplay between coherent light coupling, radiative decay and strong Rydberg-Rydberg atom interactions leads to the emergence of sizeable effective interactions while providing remarkably long coherence times. The latter are studied experimentally in a cold gas of strontium atoms for which the proposed scheme is most efficient. Our measured atom loss is in agreement with the theoretical prediction based on binary effective interactions between the driven atoms. 
\end{abstract}

\maketitle
Ultracold gases of interacting atoms provide powerful settings for exploring many-body physics in and out of equilibrium with a high level of control and experimental accessibility \cite{bdz08}. While most experiments exploit zero-range collisional interactions between atoms, the realization of finite-range interactions would open up a whole new range of study and presents an exciting frontier in the field. Among currently pursued avenues are dipolar quantum gases composed of polar molecules \cite{dgr08,nom08,mcm15} or atoms with large magnetic dipoles \cite{gwh05,by11,ksw15}, laser-cooled ion crystals \cite{lhn11,iek11,srs15,jhm15} and atoms in highly excited Rydberg states \cite{swm10}.

Owing to their large electronic orbit, Rydberg atoms feature very strong van der Waals interactions making them virtually ideal candidates for quantum simulations of spin models \cite{sce12,szf15,blr15,lbr15}. Yet, their radiative decay sets stringent limits on achievable coherence times, which thus far has prevented the observation of long-time excitation dynamics or coherent interaction effects on atomic motion. Weak off-resonant excitation of Rydberg states has been proposed \cite{mb02,hnp10,pmb10,hhp10,jr10} as a solution to this problem. Working with only a small fraction of excited atoms, this so-called Rydberg dressing promises enhanced lifetimes while providing sizeable interactions. Such interactions have recently been observed in experiments with two trapped atoms \cite{jhk15}, and are attracting broad interest in the context of quantum computing \cite{kgj13,kcm15,mge13}, frequency metrology \cite{mb02,gmb14} as well as many-body physics in quantum gases \cite{cjb10,mhs11,hcj12,dmm12,cml14,ls15,gh15} and synthetic quantum magnets \cite{gdn14,gdn15,bp15} of Rydberg-dressed atoms. Yet, the generation of sufficiently strong interactions for such applications requires large light-atom coupling strengths that are challenging to achieve, in particular for two-photon excitation of alkali atoms as employed in most current experiments.

\begin{figure}[t!]
	\includegraphics[width=\linewidth]{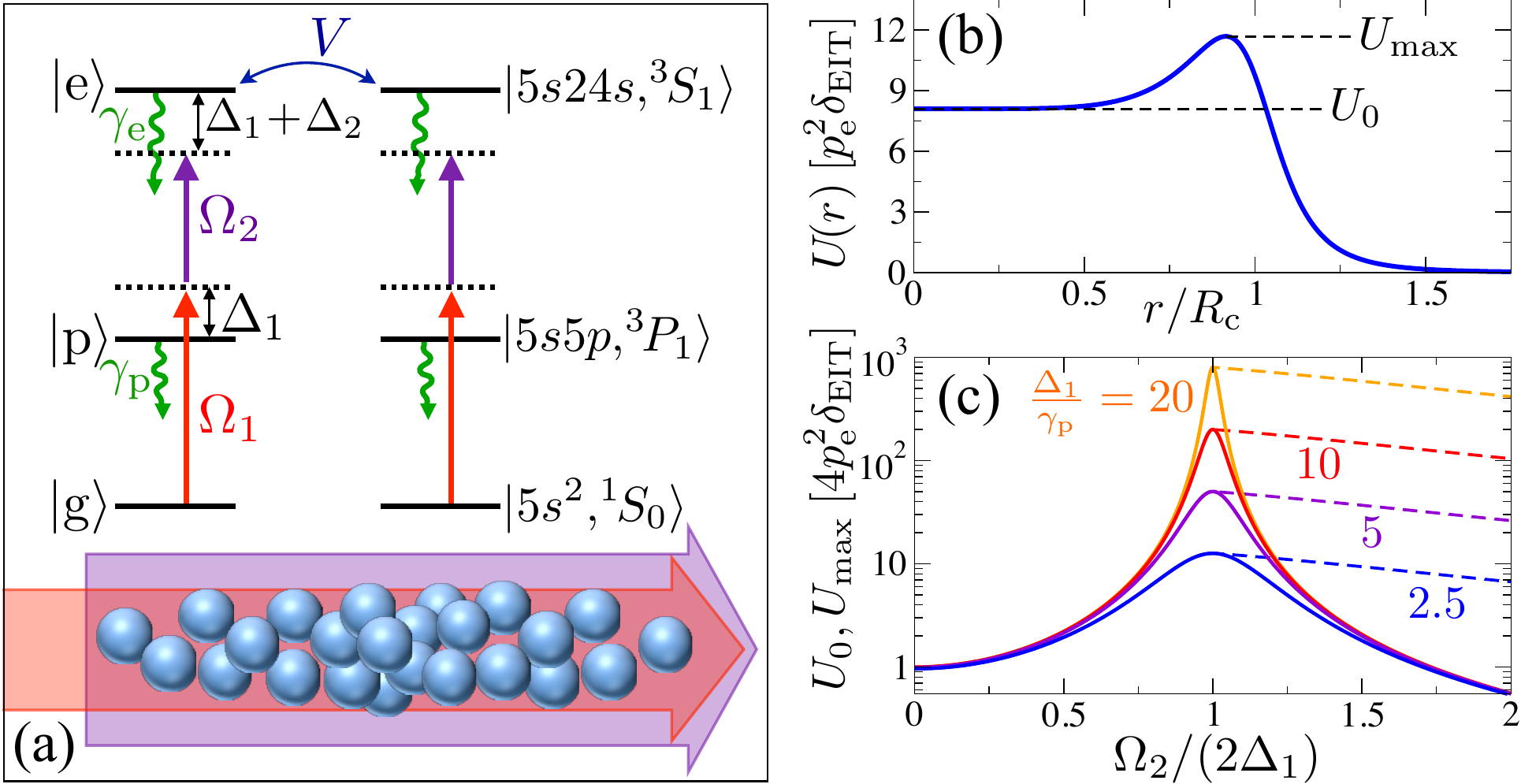}
	\caption{(a) A gas of atoms is illuminated by two laser fields that drive a two-photon transition to a Rydberg level $|{\rm e}\rangle$ with corresponding Rabi frequencies $\Omega_{1(2)}$ and detunings $\Delta_{1}$ and $\Delta_{2}$ of the lower and upper transition. The labels on the right indicate the states of Sr atoms employed in our experiments. Rydberg interactions, $V$, lead to an effective interaction potential between dressed atoms, shown in (b) for $\Delta_1/\gamma_{\rm p}=2.5$, $\Omega_2/\Delta_1=2.3$ and $\gamma_{\rm e}=0$. (c) Dependence of the characteristic energy scales $U_0$ (solid lines) and $U_{\rm max}$ (dashed lines) on $\Omega_2$ and $\gamma_{\rm p}$.
  \label{fig1}}
\end{figure}

In this work, we describe a new approach to Rydberg dressing via \emph{resonant} two-photon excitation of alkaline-earth atoms, which feature long-lived intermediate states and have recently become available to cold Rydberg gas experiments \cite{mlj10,mzs13,lbs13,dad15}. The underlying idea is based on the establishment of an approximate dark state with a strongly suppressed Rydberg state component and its modification by the mutual interactions between Rydberg atoms. We obtain a systematic solution of the $N$-body steady state in the limit of weak Rydberg excitation to show that this interplay gives rise to sizeable effective interactions between the driven atoms. In particular, we identify a two-body resonance that enables a high degree of interaction control and yields enhanced interactions at long coherence times. We experimentally realize two-photon Rydberg excitation via a long-lived triplet state of cold strontium atoms and probe decoherence due to Rydberg dressing by monitoring the induced atom loss. Our measured loss spectra reveal strong interaction effects, in quantitative agreement with the theoretical picture of emergent binary interactions between dressed ground state atoms.

We consider a gas with a density $\rho_{\rm a}$ composed of $N$ three-level atoms at positions ${\bf r}_i$, $i=1,...,N$. At low temperatures, atomic motion is much slower than typical time scales of the internal state dynamics, such that we can focus on the latter within a Born-Oppenheimer approximation \cite{hnp10}. The ground state ($|g_i\rangle$) of the $i$th atom is coupled to an intermediate state ($|p_i\rangle$) with a Rabi frequency $\Omega_1$ and frequency detuning $\Delta_1$ [cf.\ Fig.~\ref{fig1}(a)]. A second laser field drives the transition between $|p_i\rangle$ and a high lying Rydberg state ($|e_i\rangle$) with $\Omega_2$ and a total frequency detuning of $\Delta_1+\Delta_2$. The associated Hamiltonian for each atom can be written as
\begin{eqnarray}
\hat H^{(i)}&=&\frac{\Omega_1^{(i)}}{2} \left(\hat{\sigma}_{\rm gp}^{(i)} + \hat{\sigma}_{\rm pg}^{(i)}\right)+\frac{\Omega_2^{(i)}}{2} \left(\hat{\sigma}_{\rm ep}^{(i)}+\hat{\sigma}_{\rm pe}^{(i)}\right)\nonumber\\
&&-\Delta_1 \hat{\sigma}_{\rm pp}^{(i)} - (\Delta_1+\Delta_2) \hat{\sigma}_{\rm ee}^{(i)},
\end{eqnarray}
where $\hat{\sigma}_{\alpha\beta}^{(i)}=|\alpha_i\rangle\langle\beta_i|$ ($\alpha,\beta={\rm g},{\rm p}, {\rm e}$). In addition, the internal state dynamics is governed by single-particle dissipation described by Lindblad operators $\mathcal{L}_i(\hat{\rho})$ acting on the $N$-body density matrix $\hat{\rho}$ of the system. For now we include spontaneous decay of the intermediate and Rydberg state, with a rate $\gamma_{\rm p}$ and $\gamma_{\rm e}$, respectively [cf.\ Fig.~\ref{fig1}(a)]. In the absence of interactions the steady state of the corresponding master equation factorizes into one-body states $\hat{\rho}^{(1)}_i={\rm Tr}_{\bar{i}}\hat{\rho}$, where the subscript $\bar{i}$ denotes the trace over all but the $i$th atom. On two-photon resonance, $\Delta_1+\Delta_2=0$, and for negligible Rydberg state decay it assumes the particularly simple form $\hat{\rho}^{(1)}_i=|d_i\rangle\langle d_i|$ of a dark state $|d_i\rangle\propto\Omega_2|g_i\rangle-\Omega_1|e_i\rangle$ \cite{fim05}. The notion of dressing now arises in the limit $\Omega_1\ll\Omega_2$ in which the Rydberg population $p_{\rm e}=(\Omega_1/\Omega_2)^2$ can be made very small despite the resonant Rydberg-state coupling. Consequently, the rate of decoherence due to Rydberg-state decay, $p_{\rm e}\gamma_{\rm e}$ is strongly suppressed. 

For larger $p_{\rm e}$ and densities $\rho_{\rm a}$, Rydberg-Rydberg atom interactions start to become important. 
Their major effect is to shift the energy of two excited atoms, as described by the two-body operator $\hat{W}_{ij}=V({\bf r}_{ij})\hat{\sigma}_{\rm ee}^{(i)}\hat{\sigma}_{\rm ee}^{(j)}$. For Rydberg states with vanishing orbital angular momentum one can assume isotropic van der Waals interactions, $V=C_6/r_{ij}^6$, that only depend on the interatomic distance $r_{ij}=|{\bf r}_{i}-{\bf r}_j|$. The interactions of strontium atoms in $^{3\!}S_1$ Rydberg states are repulsive with van der Waals coefficients between $C_6\sim 9\mu$m$^6$MHz and $700\mu$m$^6$THz for principal quantum numbers $24<n<100$ \cite{vjp12}. Thus, the interactions can exceed typical light-matter coupling strengths, $\lesssim10$MHz, at a large distance of several $\mu$m.

\begin{figure}[t!]
	\includegraphics[width=\linewidth]{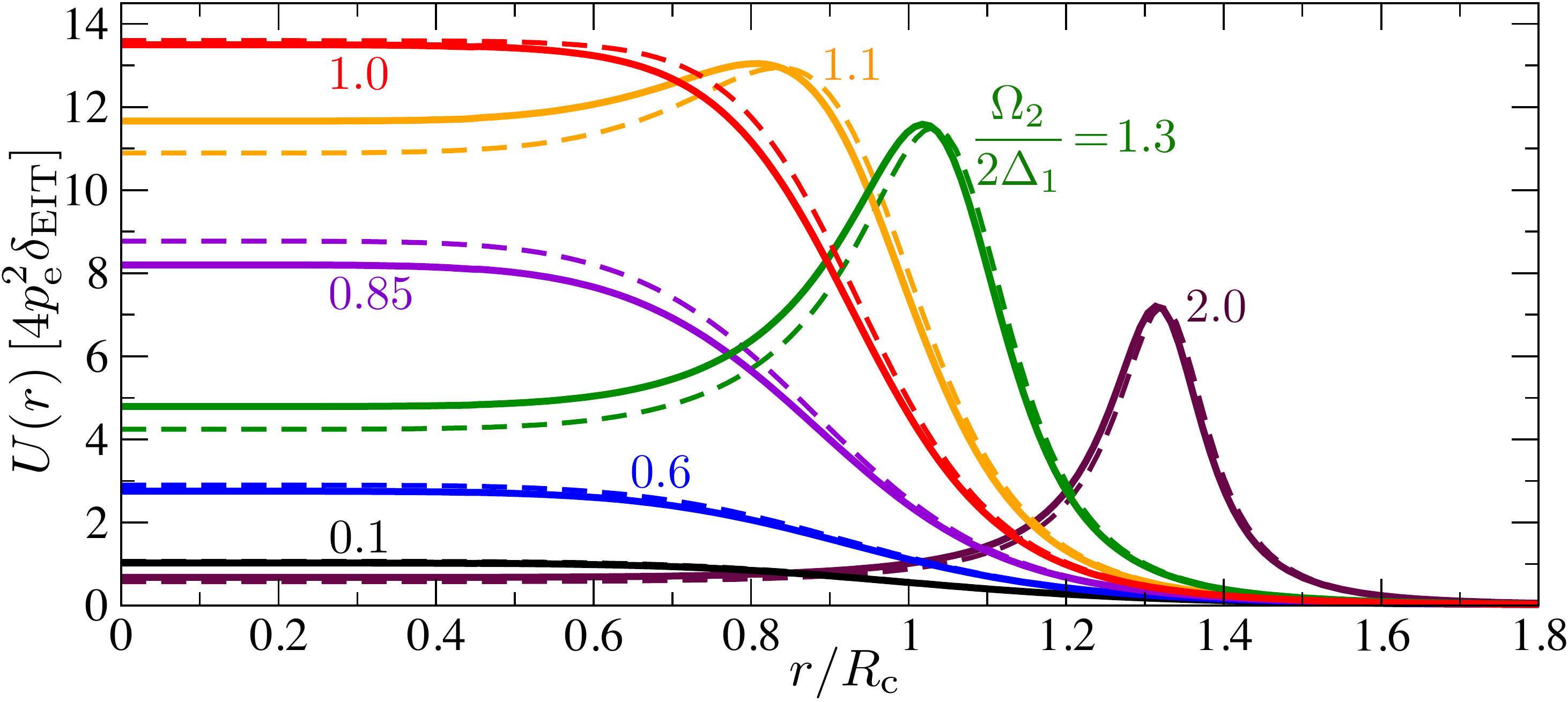}
	\caption{Effective interactions for $\Delta_1+\Delta_2=0$, $\Delta_1/\gamma_{\rm p}=2.5$, $\Omega_1/\Omega_2=0.1$, $\gamma_{\rm e}=0$  and different values $\Omega_2/(2\Delta_1)$ as indicated in the figure. The solid lines are obtained from a numerical solution of the two-body master equation and the dashed lines show the analytical expression eq.~(\ref{eq:Uanal}). \label{fig2}}
\end{figure}

While these large energy scales preclude a perturbative treatment in terms of the interaction potential $V(r)$, one can derive an approximate master equation to leading order in the small excitation fraction $p_{\rm e}$ \cite{sgh10,sha11}. To this end we first consider the dynamics of the one-body density matrix
\begin{equation}\label{eq:1body}
\partial_t\hat{\rho}^{(1)}_i=-i\left[\hat{H}_i,\hat{\rho}^{(1)}_i\right]+\mathcal{L}_i(\hat{\rho}^{(1)}_i)-i\sum_{j\neq i}{\rm Tr}_j\!\left[\hat{W}_{ij},\hat{\rho}^{(2)}_{ij}\right]
\end{equation}
which requires knowledge of the two-body matrices $\hat{\rho}^{(2)}_{ij}={\rm Tr}_{\bar{i},\bar{j}}\hat{\rho}$. Since their evolution
\begin{eqnarray}\label{eq:2body}
\partial_t\hat{\rho}^{(2)}_{ij}&=&-i\left[\hat{H}_i+\hat{H}_j,\hat{\rho}^{(2)}_{ij}\right]+\mathcal{L}_i(\hat{\rho}^{(2)}_{ij})+\mathcal{L}_j(\hat{\rho}^{(2)}_{ij})\nonumber\\
&&-i\left[\hat{W}_{ij},\hat{\rho}^{(2)}_{ij}\right]-i\sum_{k\neq i,j}{\rm Tr}_k\!\left[\hat{W}_{ik}+\hat{W}_{jk},\hat{\rho}^{(3)}_{ijk}\right],
\end{eqnarray}
in turn, requires corresponding three-body terms one eventually obtains an infinite hierarchy of equations. The last term in eq.~(\ref{eq:2body}), however, involves three excited Rydberg atoms and is effective \emph{only} for mutual distances, $r_{ik},r_{jk}\lesssim R_{\rm c}$, below a range $R_{\rm c}$ of the interaction. The Rydberg excitation number $\varepsilon=p_{\rm e}R_{\rm c}^3\rho_{\rm a}$ within the interaction volume can, thus, serve as a small parameter. Since $\hat{W}_{ij}$ projects on the doubly excited Rydberg state, the last term is smaller than the direct interaction term, $[\hat{W}_{ij},\hat{\rho}^{(2)}_{ij}]$, by a factor $\varepsilon$. To leading order in $\varepsilon$, we can, therefore, drop the three-body term and obtain a closed set of equations for the one- and two-body density matrices.

With this simplification the total energy of the system splits into a sum over binary effective interactions 
\begin{equation} \label{eq:Ueff}
U(r_{ij})={\rm Tr}[\hat{\rho}^{(2)}_{ij}(\hat{H}_i+\hat{H}_j+\hat{W}_{ij})],
\end{equation}
and the scattering rate per atom $\Gamma_i={\rm Tr}[\hat{\rho}^{(1)}_i(\gamma_{\rm p}\hat{\sigma}_{\rm pp}^{(i)}+\gamma_{\rm e}\hat{\sigma}_{\rm ee}^{(i)})]$ is also influenced by interactions with surrounding particles according to eq.~(\ref{eq:1body}).

The effective interaction, $U(r)$, can be straightforwardly obtained by numerically calculating the steady state of eq.~(\ref{eq:2body}) for a pair of atoms driven on two-photon resonance. Figure~\ref{fig2} illustrates the typical potential form for $\gamma_{\rm e}=0$. For $\Omega_2<\Delta_1$, the potential assumes a soft-core form very similar to those found for Rydberg dressing of two-level atoms \cite{hnp10,jr10}. In contrast to the latter, however, the ratio $\Omega_2/\Delta_1$ provides an additional control parameter that enables significant shaping of the potential. In particular, the interaction strength is drastically enhanced as $\Omega_2$ approaches a value of $2\Delta_1$, beyond which it develops a pronounced maximum at finite distances. As shown in Fig.~\ref{fig1}(c), the overall strength of the potential is optimized around $\Omega_2=2\Delta_1$ regardless of other laser parameters.

A better understanding of this behavior can be gained from the following analytical solution
\begin{eqnarray}\label{eq:Uanal}
&&U(r)-{\rm i} \Gamma(r) =\\
&&\frac{\Omega_1^4\, V}{\Omega_2^4}  \, 
 \frac{ 
 	4\Delta_1^2 + \gamma_{\rm p}^2 + 4 (4\Delta_1^2 + \gamma_{\rm p}^2 + \Omega_2^2) (\Delta_1-{\rm i} \gamma_{\rm p})V/\Omega_2^2
 	}
 	{ 
 	\bigl[2\Delta_1 + (4\Delta_1^2 - \Omega_2^2 - \gamma_{\rm p}^2)\frac{V}{\Omega_2^2}\bigr]^2 +  \gamma_{\rm p}^2 \bigl(1+4\frac{V \Delta_1}{\Omega_2^2}\bigr)^2
 	} \nonumber
\end{eqnarray}
which we derived in the limit $\gamma_{\rm p}\ll |\Delta_1|$, $\Omega_1\ll\Omega_2$ and $\Omega_1\ll\gamma_{\rm p}$. As shown in Fig.~\ref{fig2}, eq.~(\ref{eq:Uanal}) provides an excellent description of the numerical results, with residual deviations stemming from a violation of the last inequality for the chosen parameters.

Expectedly, the interaction scales as $p_{\rm e}^2=\Omega_1^4/\Omega_2^4$ since it requires the simultaneous Rydberg excitation of both atoms. Moreover, we can read off the potential range, $R_{\rm c}$ from the denominator of eq.~(\ref{eq:Uanal}). For $\Omega_2\ll|\Delta_1|$, $R_{\rm c}$ follows from $|V(R_{\rm c})|=\Omega_2^2/(2|\Delta_1|)$, i.e. is determined by twice the EIT linewidth $\delta_{\rm EIT}=\Omega_2^2/(4|\Delta_1|)$ as also found for effective interactions between photons in Rydberg-EIT media \cite{sha11,gof11}. In the opposite limit, $\Omega_2>|\Delta_1|$, this energy scale is replaced by $|V(\bar{R}_{\rm c})|=2|\Delta_1|$, such that 
$\bar{R}_{\rm c} = [\Omega_2/(2|\Delta_1|)]^{1/3} {R}_{\rm c}$.
Again this bears analogies to the behavior of photonic interactions in such systems \cite{pfl12,bcf14}. Likewise, eq.~(\ref{eq:Uanal}) yields an interaction strength $U(r\rightarrow0)=4p_{\rm e}^2\delta_{\rm EIT}$ for $\Omega_2\ll|\Delta_1|$, which undergoes a maximum at $\Omega=2\Delta_1$ where $U_0=U(r=0)$ is greatly enhanced by a factor $2\Delta_1^2/\gamma_{\rm p}^2$ [cf.\ Fig.~\ref{fig1}(c)]. At the same time, the corresponding scattering is suppressed by a factor $|\Delta_1|/\gamma_{\rm p}$, permitting large interactions to be obtained while ensuring long coherence times.

\begin{figure}[t!]
		\includegraphics[width=\linewidth]{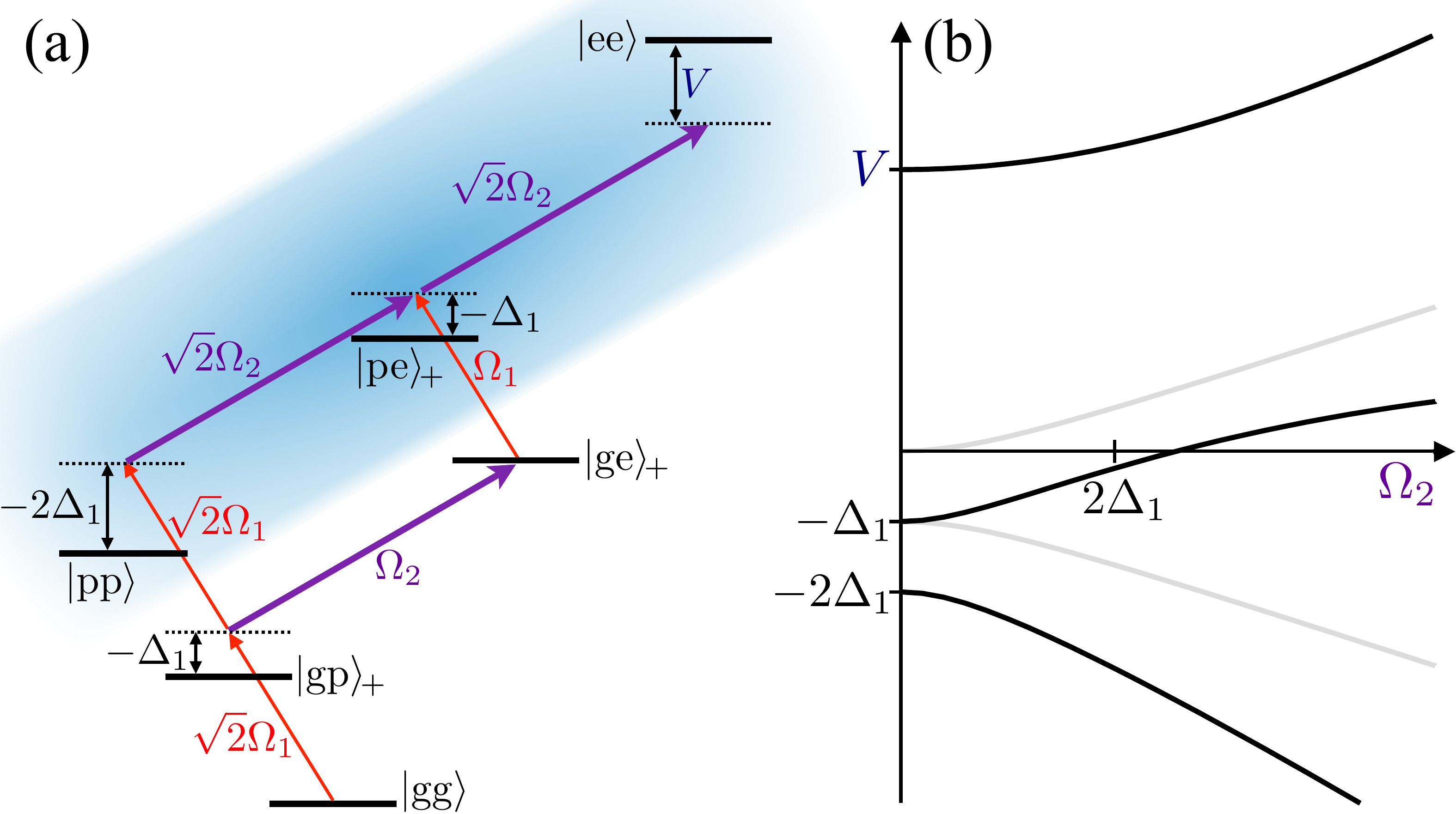}
	\caption{(a) Two-body level diagram in terms of atomic pair states on two-photon resonance ($\Delta_1+\Delta_2=0$). 
  (b) The energy spectrum of the doubly excited subspace [black lines and shaded area in (a)] indicates the occurrence of a two-body
  resonance around $\Omega_2\approx2\Delta_1$ that is driven via the singly excited states (gray lines).
  \label{fig3}}
\end{figure}

The origin of this enhancement can be traced back to a two-body resonance that emerges for strong interactions. To this end, consider the  coherent laser coupling, which can be expressed in terms of the $6$-level scheme shown in Fig.~\ref{fig3}(a), where $|\alpha\beta\rangle_+=(|\alpha\beta\rangle+|\beta\alpha\rangle)/\sqrt{2}$ denote symmetric two-particle states. For $\Omega_2\gg\Omega_1$, the system approximately separates into three subspaces composed of $|{\rm gg}\rangle$ and states with one and two excitations, respectively. Figure~\ref{fig3}(b) shows the eigenenergies of the two-excitation subspace. In the limit of large interactions the doubly excited Rydberg state decouples asymptotically and only $|{\rm pp}\rangle$ and $|{\rm pe}\rangle_+$ are hybridized by the Rydberg excitation laser. One of the two eigenenergies, $-\frac{3}{2}\Delta_1\pm\frac{1}{2}\sqrt{\Delta_1^2+2\Omega_2^2}$, vanishes at $\Omega_2=2\Delta_1$. At this point, the two-atom ground state is coupled resonantly to the corresponding eigenstate $|{\rm pp}\rangle+\sqrt{2}|{\rm pe}\rangle_+$ leading to an enhanced light shift. Hence, it is this interaction-induced two-body resonance that enables the enhanced interactions shown in Fig.~\ref{fig1}(c) and Fig.~\ref{fig2}.

\begin{figure}[b!]
		\includegraphics[width=\linewidth]{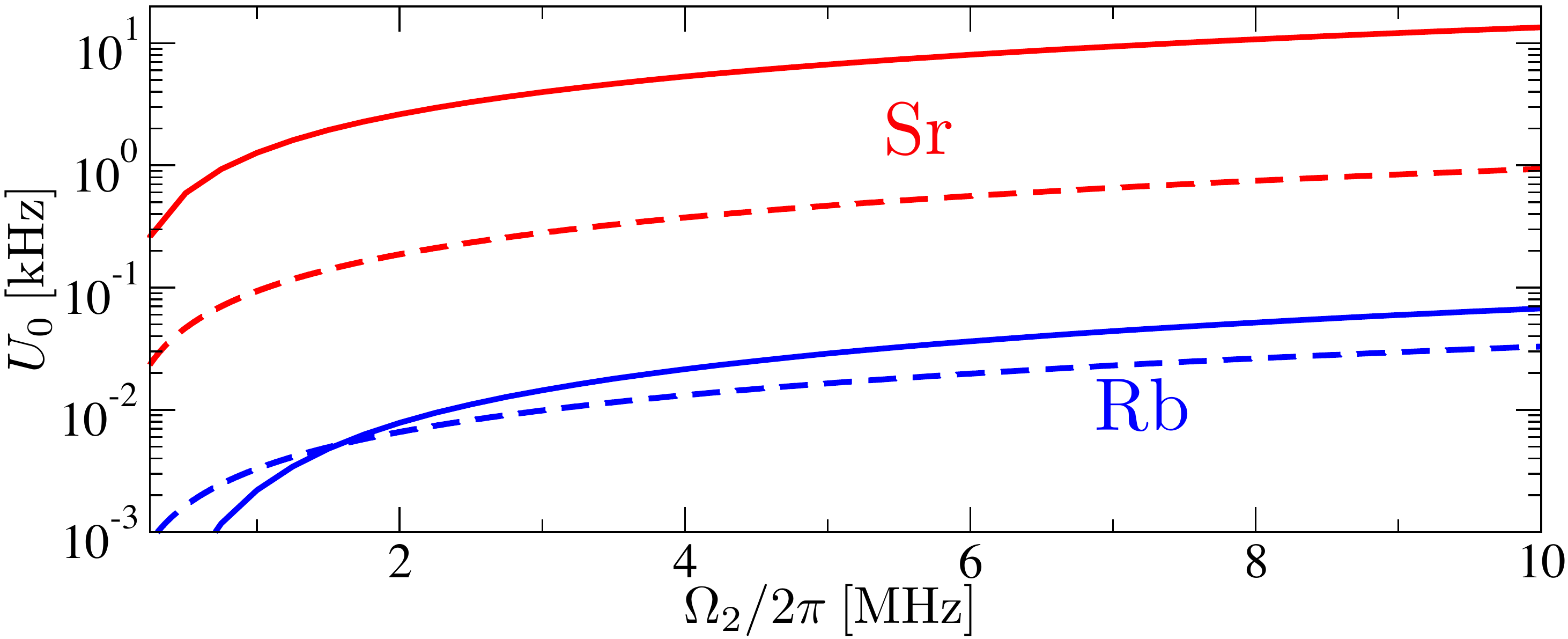}
	\caption{Maximum interaction strength at a coherence time $\Gamma^{-1}=10$ms achievable for Sr (upper two curves) and Rb (lower two curves) atoms from off-resonant dressing (dashed lines) and the approach described in this work (solid).\label{fig4}}
\end{figure}

Let us now compare the proposed approach to previous work on Rydberg dressing of two-level systems \cite{mb02,hnp10,pmb10,hhp10,jr10}. There, the underlying three-level atom is simplified to an effective two-level system, driven with a two-photon Rabi frequency $\Omega=\Omega_1\Omega_2/(2\Delta_1)$, assuming $\Omega_1\ll\Delta_1$. Optical dressing with a large two-photon detuning $\Delta_1+\Delta_2=\Delta\gg\Omega$ then yields an effective potential with strength $U_0=\Omega^4/(2\Delta)^3$ and a scattering rate of $\gamma_{\rm p}\Omega_1^2/(2\Delta_1)^2+\gamma_{\rm e}\Omega^2/(2\Delta)^2$. In Fig.~\ref{fig4} we show the interaction strength $U_0$ as a function of the Rydberg-excitation Rabi frequency $\Omega_2$ obtained from different dressing schemes for a fixed effective coherence time $\Gamma^{-1}$, while optimizing $U_0$ with respect to all remaining parameters. For strontium atoms, resonant dressing via the $(5s5p)^{3\!}P_1$ triplet state [$\gamma_{\rm p}/(2\pi)=7.6$kHz] clearly outperforms the two-level approach and yields sizeable effective interactions of several kHz, exceeding the scattering rate $\Gamma$ by up to 2 orders of magnitude. On the contrary, for alkaline atoms, such as rubidium, both approaches turn out to be rather inefficient due to the large decay rate $\gamma_{\rm p}/2\pi=6.1$MHz of the intermediate state, requiring very large $\Omega_2$ in order to realize significant effective interactions. We note that larger interactions can still be obtained from direct single-photon transitions to Rydberg $P$ states. However, the strong anisotropy of van der Waals interactions between non-zero angular momentum states \cite{sw08} can lead to pair resonances \cite{mhs11,gdn15} that cause large losses and render $P$-state dressing inapplicable in three dimensions. An important feature of our EIT-based approach is that the detrimental effects of such resonances, which can also occur at small distances \cite{jhk15,kgj13,kcm15,bp15} are completely removed. Instead of enhancing the excited Rydberg fraction an occurring pair resonance restores EIT conditions for a dressed atom pair and thereby even enhances the decoherence time instead of suppressing it (see Ref.\ \cite{suppl} for more details).

\begin{figure}
	\includegraphics[width=\linewidth]{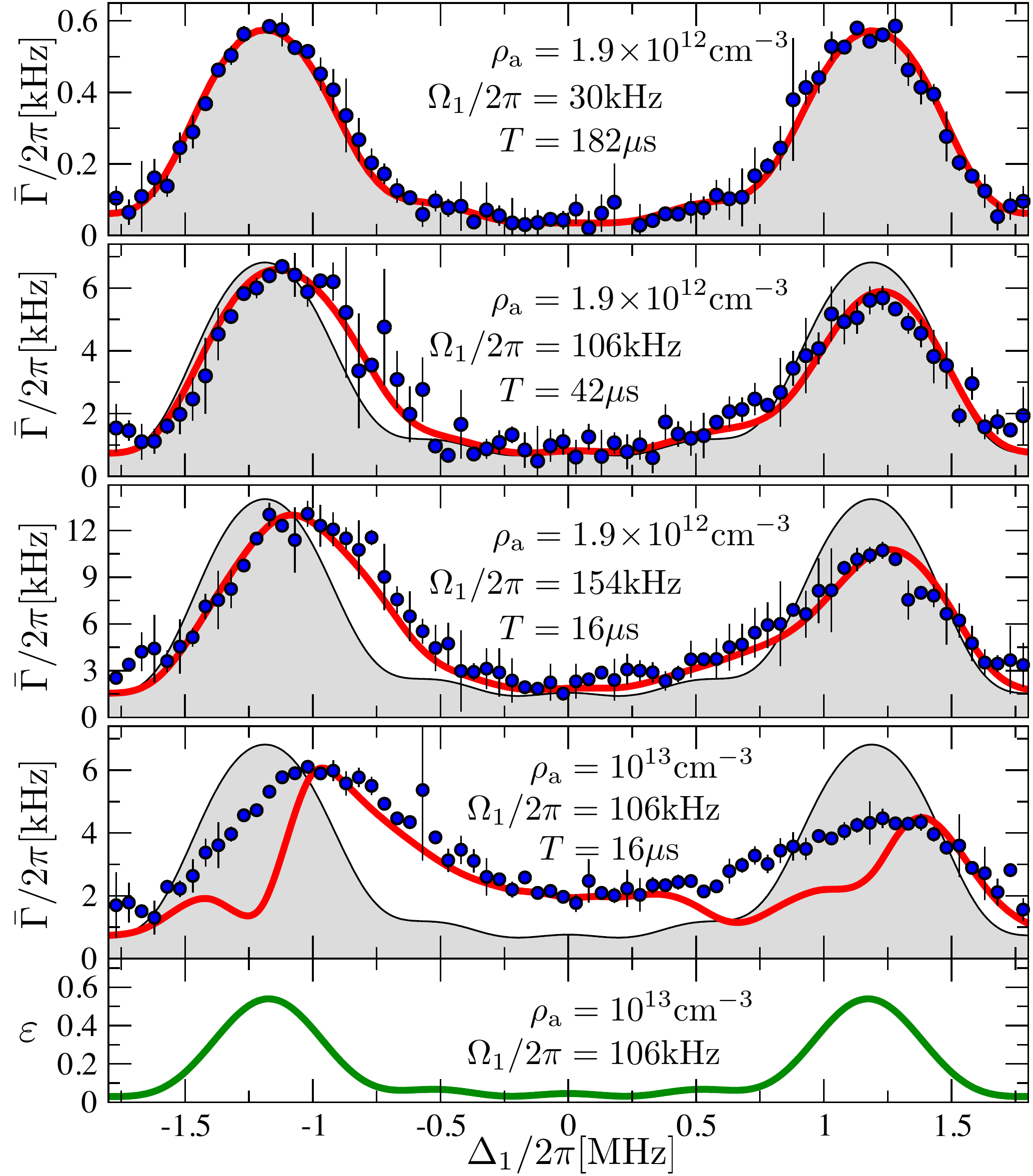}
	\vspace{-1em}
	\caption{Measured (dots) and calculated (lines) loss rates under different conditions indicated in the figure. The shaded areas show the loss spectrum without interactions. The bottom panel shows the parameter $\varepsilon$, indicating the breakdown of the leading-order theory around the Autler-Townes maxima under the indicated conditions. The theory uses $C_6/2\pi=1.4\mu$m$^6$MHz \cite{vjp12} and the remaining parameters are given in the text. \label{fig5}}
\end{figure}

We have experimentally investigated the excitation dynamics in a cold gas of $^{84}$Sr atoms \cite{suppl}. Details on the cooling and trapping are described in Ref.\ \cite{mmy09}. The atoms are excited to a $(5s24s)^{3\!}S_1$ Rydberg state via the long lived $(5s5p)^{3\!}P_1$ triplet state with a $689$nm and $319$nm laser beam with Rabi frequencies $\Omega_1$ and $\Omega_2$, respectively [cf.\ Fig.~\ref{fig1}(a)]. We probe the system by recording atom loss following laser excitation via fluorescence measurements. The gas is excited in a stroboscopic fashion with constant $\Omega_2/2\pi=2.4$MHz while driving the lower transition with a train of $\tau=2\mu$s pulses that are separated by $50\mu$s which is long compared to the Rydberg state lifetime, $\gamma_{\rm e}^{-1}\approx3.8\mu$s \cite{khk93}. The number, $T/\tau$, of pulses is chosen such that the total dressing time $T$ yields an accumulated signal well above the noise level of our detection. 

Since atom loss has been the major obstacle in previous experiments on Rydberg dressing \cite{bal14,gbb15}, we have studied losses in our measurements. As $\gamma_{\rm e}>\gamma_{\rm p}$, the observed atom loss predominantly stems from Rydberg state decay causing photon recoil and transitions to dark intermediate triplet states. Hence, we can estimate the average loss rate $\bar{\Gamma}=N_{\ell}/(NT)$ from the measured initial ($N$) and lost ($N_{\ell}$) number of atoms. 
In order to reach significant losses we chose to work on single-photon resonance ($\Delta_2=0$) and record the entire loss spectrum as a function of $\Delta_1$. Theoretically, we solve eqs.~(\ref{eq:1body}) and (\ref{eq:2body}) for a Gaussian atomic density distribution, accounting for the experimental finite laser linewidths $\gamma_1/2\pi = 21$kHz 
and $\gamma_2/2\pi=207$kHz associated with the lower and upper transition. As described above, dominant binary effective interactions emerge in the limit $\varepsilon\ll1$, in which we can neglect the three-body terms in eq.~(\ref{eq:2body}) to facilitate a numerical solution of eqs.~(\ref{eq:1body}) and (\ref{eq:2body}).
Figure~\ref{fig5} shows measured and predicted loss spectra as a function of $\Delta_1$. For our smallest density and $\Omega_1$ the measurement agrees well with the expected single-particle spectrum in the absence of interactions. Upon increasing $\Omega_1$ we observe clear density effects caused by Rydberg-Rydberg atom interactions. As shown in Fig.~\ref{fig5}, this leads to larger losses around the two-photon resonance, $\Delta_1=0$, and a growing asymmetry of the two Autler-Townes maxima around $\Delta_1=\pm\Omega_2/2$. We observe up to fourfold changes of the loss rate by the interactions, which are well described by the theory, provided $\varepsilon\ll1$. At the same time, the deviations outside of this regime indicate the emergence of higher order effective interactions between the Rydberg-dressed atoms. 

In conclusion, we have developed an approach to generate finite-range atomic interactions via two-photon resonant optical coupling to Rydberg states. This system entails interesting physics, such as an interaction-induced two-body resonance that leads to enhanced effective interactions while preserving long coherence times. Moreover, the possibility to engineer atomic interactions under Rydberg-EIT conditions \cite{pmg10} opens up new perspectives, e.g.\ concerning the coupled nonlinear dynamics of light \cite{sha11} and matter wave fields, which mutually induce effective interactions for either component based on the mechanism discussed in this work. The enhanced interactions together with the demonstrated immunity to short-distance pair resonances holds promise for studying finite-range interactions in quantum gases. Our loss measurements provide the first demonstration of strong interactions in a cold gas of triplet Rydberg states. The good agreement with the theoretical picture of dressing-induced effective interactions supports the promise of the new dressing-approach developed in this work. For longer excitation times, $\tau$, our measurements suggest the presence of an additional loss mechanism \cite{dag15} that most likely stems from dipolar interactions with $P$-state Rydberg atoms produced by black-body radiation \cite{gbb15}. The developed theoretical framework appears suitable for studying this problem and exploring potential solutions. The promising results of this work motivate future experiments to address this issue and realize narrow-bandwidth dressing off single-photon resonance ($\Delta_1\neq0$) to generate and probe coherent interactions in atomic quantum gases.\\

\begin{acknowledgments} 
We thank Rick van Bijnen for valuable discussions. This work has been supported by the EU through the FET-Open Xtrack Project HAIRS, the FET-PROACT Project RySQ, and the Marie-Curie ITN COHERENCE, by the AFOSR under Grant No.\ FA9550-14-1-0007, by the NSF under Grants No.\ 1301773 and No.\ 1205946 and by the Robert A.\ Welch Foundation under Grants No.\ C-0734 and No.\ C-1844.

\emph{Note added.}---Recently, a related article \cite{hap15} appeared. 

\end{acknowledgments}

\section{Appendix A. Experimental techniques}
Our experiment starts with a cold gas of $^{84}$Sr atoms that are confined in an optical dipole trap. The trap is formed by two crossed $1064$nm laser beams with waists of 300 $\mu$m and 440 $\mu$m in the horizontal direction and 65 $\mu$m and 38 $\mu$m in the vertical direction. The trapping beams cross at an angle of $90^{\circ}$ in the horizontal plane to form a pancake shaped trap that confines several million atoms at a temperature of $700$nK, which we further reduce to about $150$nK by evaporative cooling. 

A magnetic field of $1.5$G is applied along the vertical direction and defines the quantization axis. The $689$nm light propagates vertically upwards with a circular polarization to drive the $(5s^2)^{1\!}S_0\rightarrow(5s5p)^{3\!}P_1(m_j=+1)$ transition. We have measured the corresponding Rabi frequency from Rabi oscillation in the absence of the Rydberg coupling laser. Our UV-laser propagates horizontally and is linearly polarised in the vertical direction to drive the $(5s5p)^{3\!}P_1(m_j=+1)\rightarrow (5s24s)^{3\!}S_1 (mj=+1)$ transition. The spatial widths of the excitation beams are both much larger than the size of our cold atom cloud, such that spatial intensity variations can be neglected under the conditions of our experiments. 

The loss spectra are recorded by turning off the trapping light after Rydberg excitation and measuring the remaining number of ground state atoms via absorption imaging on the $(5s^2)^{1\!}S_0\rightarrow(5s5p)^{1\!}P_1(m_j=+1)$ at a wavelength of $461$nm.

All spectra are taken with the trap on. The trapping laser (1064\,nm) is not a magic wavelength for either the $^1S_0$--$^3P_1$ transition or the $^3P_1$--$^3S_1$ transition. 
In principle, this may lead to shifts of the resonant frequencies and a broadenenig of lines due to inhomogeneous sampling. 
For two reasons, however, these effects are negligible in our experiment. 

First, we simply take the shifted line positions as our redefined zero-detuning positions. For example,  we find the UV resonance in the trap by finding the UV laser frequency that produces symmetric Autler-Townes spectra at very low density and very low Rabi frequency for the 689\,nm laser.

Second, 
the effect of line broadening is small because the spectra are taken in such a low-intensity trap. The optical trap depth has been reduced to about 1\,$\mu$K in order to produce a 150\,nK sample. For the $^1S_0$--$^3P_1$ transition, the relevant polarizabilities are known, and the broadening of the transition is below 10\,kHz. The polarizability of the Rydberg level has not been measured, but in this optical dipole trap configuration, we observe $\approx$350\,kHz wide lines for off-resonant, two-photon excitation to the Rydberg state, which is consistent with our expected UV laser linewidth as determined by the servo-lock error signal. This is narrower than the $\approx$500\,kHz linewidth of the non-interacting Autler-Townes spectra (first panel of Fig.\ \ref{fig5}). So we conclude that the broadening of lines due to AC Stark shift is not a significant contribution to the lineshape.

\section{Appendix B. Effects of short-distance pair resonances}
Below, we provide further details on the more complex structure of the interaction potential at short interatomic distances, and show that its effects are fundamentally different in the present resonant dressing approach compared to the previously proposed off-resonant two-level scheme.

\begin{figure*}[t!]
  \includegraphics[width=0.75\textwidth]{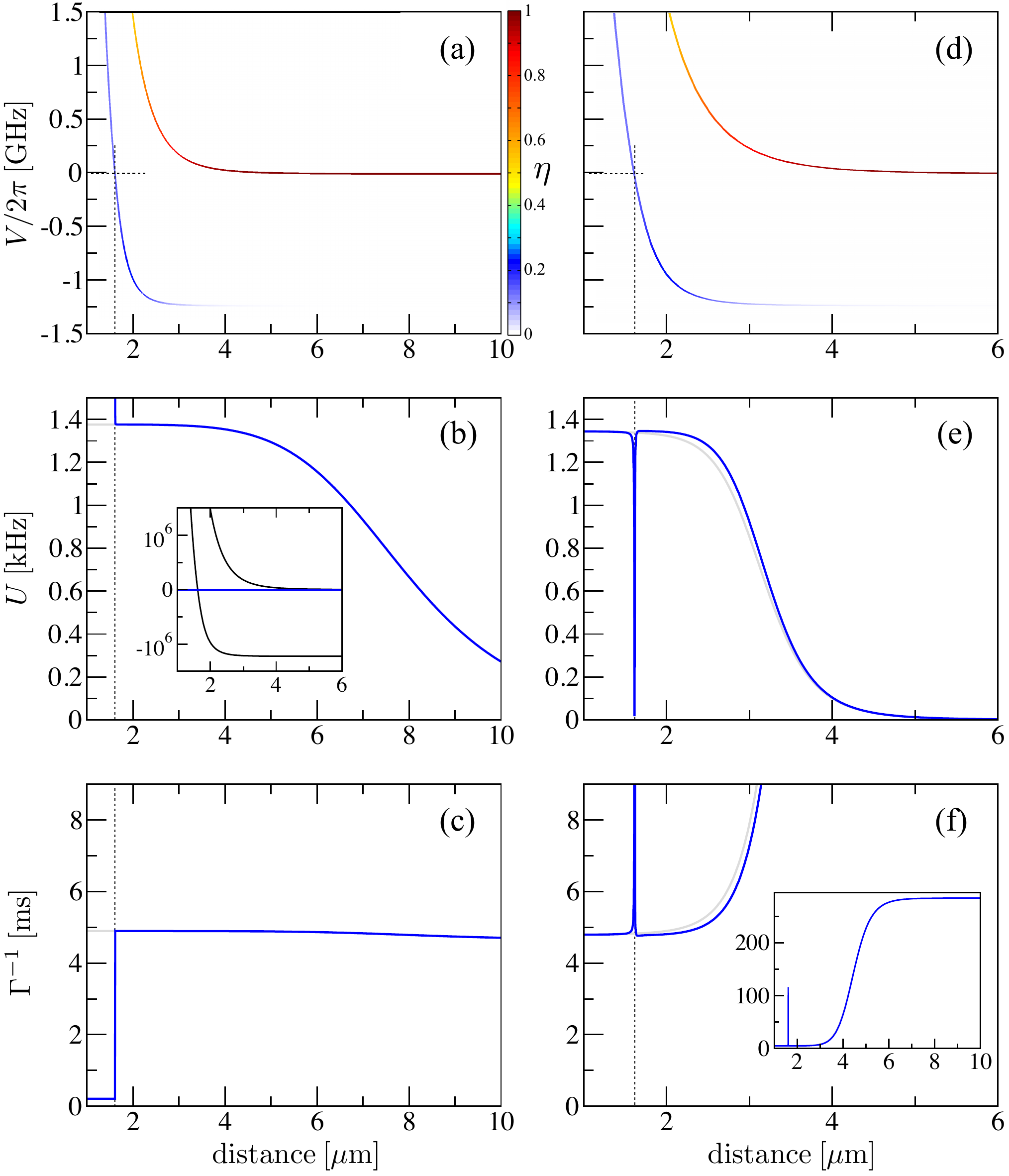}
  \caption{\label{fig:S1} (a), (d) Potential curves near the ($60S_{1/2},60S_{1/2}$)-asymptote of Rubidium atoms. The color code, $\eta=\Omega_\mu^2/\Omega_2^2$ shows the Rabi coupling $\Omega_\mu$ to a given molecular pair state relative to that of the noninteracting ($60S_{1/2},60S_{1/2}$) pair state. Panels (b) and (c) show the effective potential and lifetime resulting from off-resonant two-level dressing for $\Omega_2/2\pi=60$MHz, $\Omega_2/2\pi=0.25$MHz and $\Delta/2\pi=1.76$MHz. The grey line shows the calculation for a single potential curve with $\eta=1$ and the blue line shows the result for the two potentials of (a). The corresponding results for the resonant three-level approach are shown in (e) and (f) for $\Omega_1/2\pi=0.75$MHz, $\Omega_2/2\pi=40$MHz and $\Delta_1=\Omega_2/2$.}
\end{figure*}

While the van der Waals potential assumed in the main text provides an excellent description of the atomic interaction at large distances, the situation becomes more involved at shorter distances where the strong dipole-dipole interaction may shift further pair-states close to resonance. One can calculate such potentials by diagonalising the dipole-dipole interaction operator using a large basis of relevant pair states \cite{bp15}. This yields a set of potential curves $V_\mu(r)$ corresponding to a given molecular eigenstate $|\mu\rangle$. While there are generally many of such potential curves in the relevant energy region, only a few of them have a finite optical coupling strength $\Omega_\mu$ to the single excited pair states $|e p\rangle$ and $|pe\rangle$. To illustrate their effect on the performance of Rydberg-dressing we only include one of them [blue curve in Figs.~\ref{fig:S1}(a) and (d)] for the specific example of $60S_{1/2}$-states of Rubidium atoms.

For off-resonant two-level Rydberg-dressing such additional potential curves have a strong effect on the coherence of the system. When $V_\mu=2\Delta$ the doubly excited pair state $|\mu\rangle$ becomes resonant with the two-atom ground state $|gg\rangle$ [see inset of Fig.~\ref{fig:S1}(b)]. The resulting avoided crossing of the Rydberg dressed ground state potential and the ground state-dressed excited state \cite{kgj13} then gives rise to a drastic increase of the excited Rydberg state fraction and consequently a dramatic drop of the effective life time as shown in Fig.~\ref{fig:S1}(c). In a gas this can lead to dissipative atomic collisions and increased atom losses in the system. 

For our resonant Rydberg-dressing approach the situation is dramatically different. In this case, the resonances with other molecular pair states occur when $V_\mu=0$. However, such a resonant coupling to a doubly excited state is expected to restore EIT conditions and thereby enhance the coherence time rather than decreasing it. To address this question we numerically solve for the steady state of the driven two-atom system, now accounting for the distance dependent Rabi frequency $\Omega_\mu$ and including one additional potential curve as described above. Fig.~\ref{fig:S1}(f) demonstrates that the coherence time is indeed enhanced by the additional pair-state resonance and remains otherwise unaffected. As shown in Fig.~\ref{fig:S1}(e), this coherence time enhancement is accompanied by a sharp drop of the effective interaction potential at the resonant position. This behaviour is precisely what one would expect from the reestablishment of a two-atom dark state around the pair state resonance. Generally, the resulting potential dip is very narrow and should not affect the dynamics of colliding atom pairs compared to the approximate case without the additional potential curves [gray lines in Fig.~\ref{fig:S1}(e)].


\begin{thebibliography}{99}
\bibitem{bdz08} I. Bloch, J. Dalibard and W. Zwerger, Rev. Mod. Phys. {\bf 80}, 885 (2008).
\bibitem{dgr08} J. Deiglmayr {\it et al.}, Phys. Rev. Lett. {\bf 101}, 133004 (2008).
\bibitem{nom08} K.-K. Ni {\it et al.}, Science {\bf 322}, 231 (2008).
\bibitem{mcm15} S.\,A. Moses {\it et al.}, Science {\bf 350}, 659 (2015).
\bibitem{gwh05} A. Griesmaier {\it et al.}, Phys. Rev. Lett. {\bf 94} 160401 (2005).
\bibitem{by11} M. Lu {\it et al.}, Phys. Rev. Lett. {\bf 107} 190401 (2011).
\bibitem{ksw15} H. Kadau {\it et al.}, \href{http://dx.doi.org/10.1038/nature16485}{Nature {\bf 530}, 194 (2016)}.  
\bibitem{lhn11} B.\,P. Lanyon {\it et al.}, Science {\bf 334}, 57 (2011).
\bibitem{iek11} R. Islam {\it et al.}, Nature Commun. {\bf 2}, 377 (2011).
\bibitem{srs15} C. Senko {\it et al.}, Phys. Rev. X {\bf 5}, 021026 (2015).
\bibitem{jhm15} P. Jurcevic {\it et al.}, Phys. Rev. Lett. {\bf 115}, 100501 (2015).
\bibitem{swm10} M. Saffman, T. G. Walker and K. M\o{}lmer, Rev. Mod. Phys. {\bf 82}, 2313 (2010).
\bibitem{sce12} P. Schau\ss\  {\it et al.}, Nature {\bf 491}, 87 (2012).
\bibitem{szf15} P. Schau\ss\  {\it et al.}, Science {\bf 347}, 1455 (2015).
\bibitem{blr15} D. Barredo {\it et al.}, Phys. Rev. Lett. {\bf 114}, 113002 (2015).
\bibitem{lbr15} H. Labuhn {\it et al.}, arXiv:1509.04543
\bibitem{mb02} I. Bouchoule and K. M\o{}lmer, Phys. Rev. A {\bf 65}, 041803(R) (2002).
\bibitem{hnp10} N. Henkel, R. Nath and T. Pohl, Phys. Rev. Lett. {\bf 104}, 195302 (2010).
\bibitem{pmb10} G. Pupillo {\it et al.}, Phys. Rev. Lett. {\bf 104}, 223002 (2010).
\bibitem{hhp10} J. Honer {\it et al.}, Phys. Rev. Lett. {\bf 105}, 160404 (2010).
\bibitem{jr10} J.\,E. Johnson and S. L. Rolston, Phys. Rev. A {\bf 82}, 033412 (2010).
\bibitem{jhk15} Y.-Y. Jau {\it et al.}, \href{http://dx.doi.org/10.1038/nphys3487}{Nature Phys. {\bf 12}, 71 (2016)}.
\bibitem{kgj13} T. Keating {\it et al.}, Phys. Rev. A {\bf 87}, 052314 (2013).
\bibitem{kcm15} T. Keating {\it et al.}, Phys. Rev. A {\bf 91}, 012337 (2015).
\bibitem{mge13} S. M\"obius {\it et al.}, Phys. Rev. A {\bf 87}, 051602(R) (2013).
\bibitem{gmb14} L.\,I.\,R. Gil {\it et al.}, Phys. Rev. Lett. {\bf 112}, 103601 (2014).
\bibitem{cjb10} F. Cinti {\it et al.}, Phys.\ Rev.\ Lett.\ {\bf 105}, 135301 (2010).
\bibitem{mhs11} F. Maucher {\it et al.}, Phys.\ Rev.\ Lett.\ {\bf 106}, 170401 (2011).
\bibitem{hcj12} N. Henkel {\it et al.}, Phys. Rev. Lett. {\bf 108}, 265301 (2012).
\bibitem{dmm12} A. Dauphin, M. M\"uller, and M. A. Martin-Delgado, Phys. Rev. A {\bf 86}, 053618 (2012).
\bibitem{cml14} F. Cinti {\it et al.}, Nature Commun. {\bf 5}, 3235 (2014).
\bibitem{ls15} X. Li and S. Das Sarma, Nature Commun. {\bf 6}, 7137 (2015).
\bibitem{gh15} A. Gei\ss ler, I. Vasic, and W. Hofstetter, arXiv:1509.06292
\bibitem{gdn14} A.\,W. Glaetzle {\it et al.}, Phys. Rev. X {\bf 4}, 041037 (2014).
\bibitem{gdn15} A.\,W. Glaetzle {\it et al.}, Phys. Rev. Lett. {\bf 114}, 173002 (2015).
\bibitem{bp15} R.\,M.\,W. van Bijnen and T. Pohl, Phys. Rev. Lett. {\bf 114}, 243002 (2015).
\bibitem{mlj10} J. Millen, G. Lochead, and M.\,P.\,A. Jones, Phys. Rev. Lett. {\bf 105}, 213004 (2010).
\bibitem{mzs13} P. McQuillen {\it et al.}, Phys. Rev. A {\bf 87}, 013407 (2013).
\bibitem{lbs13} G. Lochead {\it et al.}, Phys. Rev. A {\bf 87}, 053409 (2013).
\bibitem{dad15} B.\,J. DeSalvo {\it et al.}, Phys. Rev. A {\bf 92}, 031403 (2015).
\bibitem{fim05} M. Fleischhauer, A. Imamoglu, and J.\,P. Marangos, Rev. Mod. Phys. {\bf 77}, 633 (2005).
\bibitem{vjp12} C.\,L. Vaillant, M.\,P.\,A. Jones, and R.\,M. Potvliege, J. Phys. B {\bf 45} 135004 (2012).
\bibitem{sgh10} H. Schempp {\it et al.}, Phys. Rev. Lett. {\bf 104}, 173602 (2010).
\bibitem{sha11} S. Sevincli {\it et al.}, Phys. Rev. Lett. {\bf 107}, 153001 (2011).
\bibitem{gof11} A.\,V. Gorshkov {\it et al.}, Phys. Rev. Lett. {\bf 107}, 133602 (2011).
\bibitem{pfl12} T. Peyronel {\it et al.}, Nature {\bf 488}, 57 (2012).
\bibitem{bcf14} P. Bienias {\it et al.}, Phys. Rev. A {\bf 90}, 053804 (2014).
\bibitem{sw08} T.\,G. Walker and M. Saffman, Phys. Rev. A {\bf 77}, 032723 (2008).
\bibitem{suppl} See Appendix A for further details of our experiments and Appendix B for more details and explicit calculations on the effects of additional pair resonances.
\bibitem{mmy09} Y.\,N. Martinez de Escobar {\it et al.}, Phys. Rev. Lett. {\bf 103}, 200402 (2009).
\bibitem{khk93} S. Kunze {\it et al.}, Z. Phys. D {\bf 27}, 111 (1993).
\bibitem{bal14} J.\,B. Balewski {\it et al.}, New J. Phys. {\bf 16}, 063012 (2014).
\bibitem{gbb15} E.\,A. Goldschmidt {\it et al.}, 
\href{http://dx.doi.org/10.1103/PhysRevLett.116.113001}{Phys.\ Rev.\ Lett.\ {\bf 116}, 113001 (2016)}.
\bibitem{pmg10} J.\,D. Pritchard {\it et al.}, Phys. Rev. Lett. {\bf 105}, 193603 (2010).
\bibitem{dag15} B.\,J. DeSalvo {\it et al.}, \href{http://dx.doi.org/10.1103/PhysRevA.93.022709}{Phys.\ Rev.\ A {\bf 93}, 022709 (2016)}.
\bibitem{hap15} S. Helmrich {\it et al.}, \href{http://dx.doi.org/10.1088/0953-4075/49/3/03LT02}{J.\ Phys.\ B. {\bf 49}, 03LT02 (2016)}.
\end{thebibliography}
\end{document}